\documentclass[pra,twocolumn,showpacs,superscriptaddress,10pt]{revtex4-1}
\usepackage[colorlinks,citecolor=blue,linkcolor=red,dvipdfm]{hyperref}
\usepackage{amsmath,amssymb,graphicx}

\begin{document}

\title{Stable and oscillating solitons of $\mathcal{PT}$-symmetric couplers with gain and loss in fractional dimension}

\author{Liangwei Zeng}
\affiliation{College of Physics and Optoelectronic Engineering, Shenzhen University, Shenzhen 518060, China}
\affiliation{Shenzhen Key Laboratory of Micro-Nano Photonic Information Technology,
College of Physics and Optoelectronic Engineering, Shenzhen University, Shenzhen 518060, China}

\author{Jincheng Shi}
\affiliation{State Key Laboratory of Transient Optics and Photonics, Xi'an
Institute of Optics and Precision Mechanics of Chinese Academy of Sciences, Xi'an 710119, China}

\author{Xiaowei Lu}
\email{\underline{xiaoweilu@szu.edu.cn}}
\affiliation{College of Physics and Optoelectronic Engineering, Shenzhen University, Shenzhen 518060, China}
\affiliation{Shenzhen Key Laboratory of Micro-Nano Photonic Information Technology,
College of Physics and Optoelectronic Engineering, Shenzhen University, Shenzhen 518060, China}

\author{Yi Cai}
\affiliation{College of Physics and Optoelectronic Engineering, Shenzhen University, Shenzhen 518060, China}
\affiliation{Shenzhen Key Laboratory of Micro-Nano Photonic Information Technology,
College of Physics and Optoelectronic Engineering, Shenzhen University, Shenzhen 518060, China}

\author{Qifan Zhu}
\affiliation{College of Physics and Optoelectronic Engineering, Shenzhen University, Shenzhen 518060, China}
\affiliation{Shenzhen Key Laboratory of Micro-Nano Photonic Information Technology,
College of Physics and Optoelectronic Engineering, Shenzhen University, Shenzhen 518060, China}

\author{Hongyi Chen}
\affiliation{College of Physics and Optoelectronic Engineering, Shenzhen University, Shenzhen 518060, China}
\affiliation{Shenzhen Key Laboratory of Micro-Nano Photonic Information Technology,
College of Physics and Optoelectronic Engineering, Shenzhen University, Shenzhen 518060, China}

\author{Hu Long}
\affiliation{College of Physics and Optoelectronic Engineering, Shenzhen University, Shenzhen 518060, China}
\affiliation{Shenzhen Key Laboratory of Micro-Nano Photonic Information Technology,
College of Physics and Optoelectronic Engineering, Shenzhen University, Shenzhen 518060, China}

\author{Jingzhen Li}
\email{\underline{lijz@szu.edu.cn}}
\affiliation{College of Physics and Optoelectronic Engineering, Shenzhen University, Shenzhen 518060, China}
\affiliation{Shenzhen Key Laboratory of Micro-Nano Photonic Information Technology,
College of Physics and Optoelectronic Engineering, Shenzhen University, Shenzhen 518060, China}

\begin{abstract}
Families of coupled solitons of $\mathcal{PT}$-symmetric physical models with gain and loss in fractional dimension and in settings with and without cross-interactions modulation (CIM), are reported. Profiles, powers, stability areas, and propagation dynamics of the obtained $\mathcal{PT}$-symmetric coupled solitons are investigated. By comparing the results of the models with and without CIM, we find that the stability area of the model with CIM is much broader than the one without CIM. Remarkably, oscillating $\mathcal{PT}$-symmetric coupled solitons can also exist in the model of CIM with the same coefficients of the self- and cross-interactions modulations. In addition, the period of these oscillating coupled solitons can be controlled by the linear coupling coefficient.
\end{abstract}

\maketitle

\section{Introduction}
The nonlinear Schr\"{o}dinger equation (NLSE) \cite{NLSE1,NLSE2,NLSE3,NLSE4,NLSE5,NLSE6,NLSE7,NLSE8,NLSE8b,NLSE8c}, which can be used to describe the dynamics of optical solitons in laser beam or  matter-wave solitons in Bose-Einstein condensate (BEC), has drawn more and more attention in recent decades. It is also widely used in quantum mechanics \cite{NLSE9}, hydrodynamics \cite{NLSE10}, nonlinear optics \cite{NLSE3,NLSE7}, BEC \cite{NLSE3,NLSE7} as well as superconductivity \cite{NLSE11}. Note that it is easy to stabilize a one-dimensional soliton, since it is now widely known that the bright soliton is the exact solution of one-dimensional NLSE with uniform cubic self-focusing nonlinearity. However, the collapse will happen in two-dimensional NLSE with uniform cubic self-focusing nonlinearity \cite{COLLAPSE}. A popular way to avoid such collapse is the introduction of the linear potentials \cite{LP1,LP2,LP3}, which can stabilize various kinds of solitons in all dimensions. Another way to create stable solitons in high dimensions is the introduction of nonlinearities, such as the employment of periodically modulated nonlinearities \cite{NLSE3,MN1,MN2} or spatially inhomogeneous defocusing nonlinearities \cite{DEF1,DEF2,DEF3,DEF4,DEF5}.

It is worth to mention that single NLSE can only describe the propagation of an optical soliton in a laser beam or the evolution of a matter-wave soliton in a BEC. As for the settings of two (or more) laser beams or BECs, the models of coupled NLSEs \cite{CNLSE1,CNLSE2,CNLSE3,CNLSE4,CNLSE5,CNLSE6,CNLSE7} have to be considered due to the coupled interactions of the laser beams or BECs. These models are widely studied in optical couplers, especially in the fields of optical fiber \cite{CNLSE8,CNLSE9} and waveguide \cite{CNLSE10} in recent years. Further, the models of couplers with parity-time- ($\mathcal{PT}$-) symmetry have drawn more and more attention, since Bender demonstrated that non-Hermitian Hamiltonians with $\mathcal{PT}$-symmetry can have real spectra \cite{PT0}, which is an important extension of standard quantum mechanics. These physical models with $\mathcal{PT}$-symmetry exhibit some novel physical properties due to their non-Hermitian Hamiltonians \cite{PT1,PT2} and various physical settings with $\mathcal{PT}$-symmetry have been widely investigated during the past decade \cite{PT3,PT4,PT5,PT6,PT7,PT8,PT9}.

The fractional derivative \cite{Frac}, firstly introduced by Laskin, is another great extension of the standard quantum mechanics \cite{Lask1,Lask2,Lask3}. Under such condition, the quantum mechanics becomes the fractional quantum mechanics, and the NLSE becomes the nonlinear fractional Schr\"{o}dinger equation (NLFSE). Followed by the above landmark study of Laskin in the NLSE, various studies have been reported in recent years \cite{Frac00,Frac0,Frac1,Frac2,Frac3,Frac4,Frac5,Frac6,Frac7,Frac8,Frac9,Frac10,Frac11,Frac12,Frac13,Frac14,Frac15,Frac15b,Frac16,Frac17,Frac18,Frac19,Frac19b,Frac19c,Frac20,Frac21,Frac22,Frac23,Frac24,Frac25,Frac26,Frac27,Frac28}, including the propagation of light beams \cite{Frac1,Frac2}, accessible solitons \cite{Frac5,Frac6}, gap solitons \cite{Frac7,Frac16}, vortex solitons \cite{Frac18,Frac19c,Frac27}, solitons in nonlinear lattices \cite{Frac14}, and soliton clusters \cite{Frac19,Frac19b}.

Despite so many researches that have been reported in the NLFSE, the coupled $\mathcal{PT}$-symmetric NLFSEs with gain and loss, which can be used to describe the propagations of binary $\mathcal{PT}$-symmetric laser beams or BECs with coupled interactions, have not been yet reported, to the best of our knowledge. This coupled model can be employed to study the optical waveguides and fiber couplers, which have great potential in optical communications. The objective of this work is to extend the NLFSE to its coupled form with $\mathcal{PT}$-symmetry, from which we can explore the existence, stability and other properties of optical and matter-wave solitons in this model under fractional-order diffraction.

This paper is organized as follows. In Sec. \ref{sec2}, we introduce the theoretical model and report some typical analytical solutions for coupled solitons. Sec. \ref{sec3}, which presents numerical results for $\mathcal{PT}$-symmetric coupled solitons, is divided into two parts. The numerical results for the model without cross-interactions modulations are given in Sec. \ref{sec3a}, and the corresponding numerical results for the model with cross-interactions modulations are reported in Sec. \ref{sec3b}. We find that the field profiles and stability domains of coupled solitons are strongly affected by the linear coupling coefficient, the fractional-order diffraction described by the L\'{e}vy index $\alpha$, the propagation constant, and the cross-interactions modulation. It should be mentioned that these coupled solitons can exist only when the gain/loss parameter ($\gamma$) is less than the linear coupling coefficient ($\kappa$). Finally, the paper is summarized in Sec. \ref{sec4}.

\section{Theoretical model}
\label{sec2}
The propagations of laser beams in $\mathcal{PT}$-symmetric couplers with gain and loss under fractional-order diffraction can be described by the dimensionless coupled NLFSEs
\begin{equation}
\left\{
\begin{aligned}
i\frac{\partial U_1}{\partial z}=&\frac{1}{2}\left(-\frac{\partial^2}{\partial x^2}\right)^{\alpha/2}U_1-\mathrm{g}\left|U_1\right|^2U_1-\epsilon\left|U_2\right|^2U_1 \\
&-\kappa U_2+i\gamma U_1, \\
i\frac{\partial U_2}{\partial z}=&\frac{1}{2}\left(-\frac{\partial^2}{\partial x^2}\right)^{\alpha/2}U_2-\mathrm{g}\left|U_2\right|^2U_2-\epsilon\left|U_1\right|^2U_2 \\
&-\kappa U_1-i\gamma U_2.
\end{aligned}
\label{NLFSE}
\right.
\end{equation}
Here $U_{1,2}$ and $z$ represent the field amplitudes and propagation distance, respectively, $\mathrm{g}>0$ and $\epsilon>0$ represent the coefficients of attractive self- and cross-interactions modulations of the components, respectively. Note that $\mathrm{g}<0$ and $\epsilon<0$ denote the coefficients of repulsive self- and cross-interactions modulations of the components that can be used to generate dark solitons \cite{DARK}. In this article, we  focus on the formation of bright solitons, hence the case when $\mathrm{g}<0$ and $\epsilon<0$ is not discussed here. The parameter $\kappa>0$ denotes the linear coupling  coefficient, and $\gamma>0$, that is the coefficient of gain and loss, stands for the gain in one component and the loss in the other one.   Note that $(-\partial^2/\partial x^2)^{\alpha/2}$ denotes the fractional derivative, where $\alpha$ ($1<\alpha\leq2$) stands for the L\'{e}vy index. The definition of fractional derivative is as follows \cite{Lask1,Lask2,Lask3}
\begin{equation}
\begin{split}
\left( -\frac{\partial ^{2}}{\partial x^{2}}\right) ^{\alpha /2}U=&\frac{1}{2\pi }\int_{-\infty }^{+\infty }|s|^{\alpha }ds \\
\times & \int_{-\infty }^{+\infty}d\zeta \exp \left[ ip\left( x-\zeta \right) \right] U(\zeta ).
\end{split}
\end{equation}
In particular, Eqs. (\ref{NLFSE}) will degenerate to the generic $\mathcal{PT}$-symmetric coupled nonlinear Schr\"{o}dinger equations when $\alpha=2$. As for BECs, $U_{1,2}$ and $z$ should be replaced by the wave functions $\phi_{1,2}$ and time $t$.

\begin{figure}[tbp]
\begin{center}
\includegraphics[width=1\columnwidth]{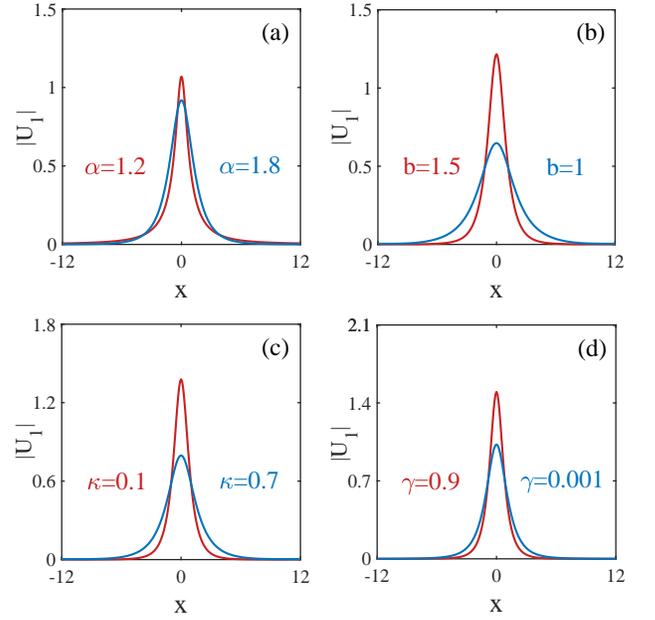}
\end{center}
\caption{Profiles of solitons (shown for components of $U_1$): (a) with different values of $\alpha$ at $\kappa=0.8$, $\gamma=0.001$, $b=1.2$; (b) with different values of $b$ at $\alpha=1.8$, $\kappa=0.8$, $\gamma=0.001$; (c) with different values of $\kappa$ at $\alpha=1.8$, $\gamma=0.001$, $b=1$; (d) with different values of $\gamma$ at $\alpha=1.8$, $\kappa=1$, $b=1.5$. The profiles of $|U_2|$ are similar to $|U_1|$. $\epsilon=0$ are used in Figs. \ref{fig1}$\sim$\ref{fig3}. $\mathrm{g}=1$ are used throughout this paper.}
\label{fig1}
\end{figure}

With the real propagation constant $b$ ($b$ should be replaced by chemical potential $-\mu$ in BECs), the stationary solutions of these $\mathcal{PT}$-symmetric coupled solitons with gain and loss can be found by generic forms
\begin{equation}
\left\{
\begin{aligned}
U_1(x,z)=U(x){\rm exp}(ibz-i\delta/2),\\
U_2(x,z)=U(x){\rm exp}(ibz+i\delta/2).
\end{aligned}
\label{NLFSES}
\right.
\end{equation}
Here $\delta$ denotes the constant phase shift between the components, which is given by
\begin{equation}
\delta={\rm arcsin}(\gamma/\kappa).
\label{CPS}
\end{equation}
As is seen from Eqs. (\ref{NLFSES}) and (\ref{CPS}), the difference between the two components $U_{1,2}$, which is caused by their parameter of gain-loss, is their imaginary part ${\rm exp}(\pm i\delta/2)$. Due to their opposite parameter of gain-loss, these two components will have the opposite transition of energy in their propagations, which can be referred to the following Figs. \ref{fig3}, \ref{fig6}, \ref{fig7}, \ref{fig8} and \ref{fig9}.

Here the real function $U(x)$  satisfies the generic stationary NLFSE with a varying value of the propagation constant $b$
\begin{equation}
-(b-\kappa_s)U=\frac{1}{2}\left(-\frac{\partial^2}{\partial x^2}\right)^{\alpha/2}U-\mathrm{g} U^3-\epsilon U^3,
\label{NLFSE2}
\end{equation}
\begin{equation}
\kappa_s=\sqrt{\kappa^2-\gamma^2}.
\label{KS}
\end{equation}
According to Eq. (\ref{KS}), it is easy to see that these coupled solitons can exist only when
\begin{equation}
\gamma\leq\gamma_{max}=\kappa.
\label{GAMMAK}
\end{equation}
From Eq. (\ref{GAMMAK}), the largest value of the gain-loss coefficient in these $\mathcal{PT}$-symmetric couplers is limited to the value of the linear coupling coefficient $\kappa$.

\begin{figure}[tbp]
\begin{center}
\includegraphics[width=1\columnwidth]{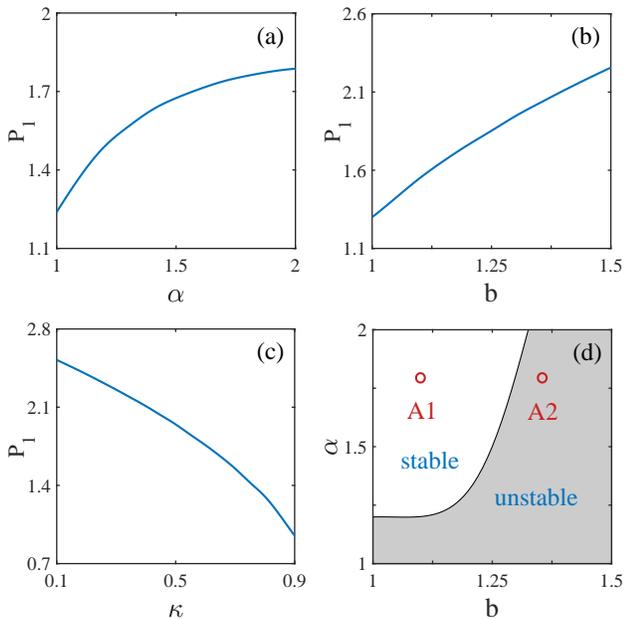}
\end{center}
\caption{(a) Soliton power $P$ versus $\alpha$ at $\kappa=0.8$, $\gamma=0.001$, $b=1.2$. (b) $P$ versus $b$ at $\alpha=1.8$, $\kappa=0.8$, $\gamma=0.001$. (c) $P$ versus $\kappa$ at $\alpha=1.8$, $\gamma=0.001$, $b=1$. (d) Stability (white) and instability (gray) domains for the coupled solitons with $\kappa=0.8$, $\gamma=0.001$ in the ($b,\alpha$) plane. The propagations of the solitons corresponding to the points marked by A1 and A2 are shown in Figs. \ref{fig3}(a,b) and \ref{fig3}(c,d), respectively.}
\label{fig2}
\end{figure}

\begin{figure}[tbp]
\begin{center}
\includegraphics[width=1\columnwidth]{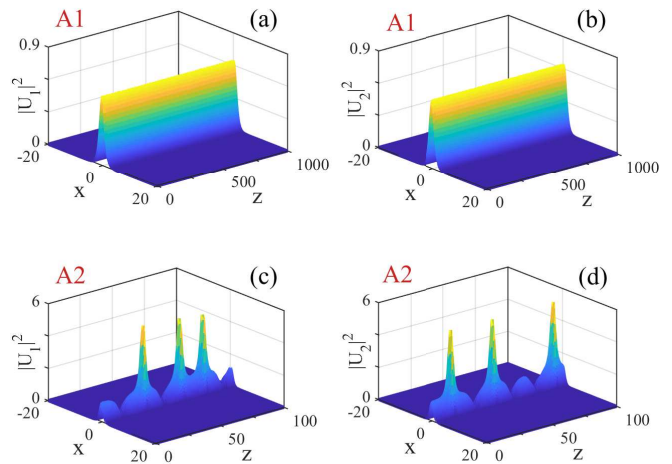}
\end{center}
\caption{Propagations of coupled solitons with different values of $b$ at $\alpha=1.8$, $\kappa=0.8$, $\gamma=0.001$: (a,b) with $b=1$; (c,d) with $b=1.35$. The left~/~right column displays the $U_1$~/~$U_2$ component.}
\label{fig3}
\end{figure}

Particularly, analytical solution of Eq. (\ref{NLFSE2}) can be solved when $\alpha=2$, which is given by
\begin{equation}
U=\sqrt{2(b-\kappa_s)/(\mathrm{g}+\epsilon)}{\rm sech}\left(\sqrt{2(b-\kappa_s)}x\right),
\label{SOLUTION}
\end{equation}
where $\kappa_s$ is defined by Eq. (\ref{KS}).

The soliton power $P$ of $U_{1,2}$ is defined by $P_1=\int\left|U_1\right|^2dx$, $P_2=\int\left|U_2\right|^2dx$. In this paper, the modified squared-operator method \cite{MSOM} and split-step Fourier method are employed to solve the stationary solutions for Eq. (\ref{NLFSE2}) and numerical simulations for Eqs. (\ref{NLFSE}) respectively.

\section{Numerical results}
\label{sec3}

In this section, we focus on presenting our numerical results of $\mathcal{PT}$-symmetric coupled solitons with gain and loss under different parameters in the models of Eqs. (\ref{NLFSE}) and Eq. (\ref{NLFSE2}), inlcuding the $\mathcal{PT}$-symmetric coupled solitons with different linear coupling coefficients, fractional-order diffractions, propagation constants and cross-interactions modulations. Particularly, we find that the cross-interactions modulation can greatly affect these coupled solitons. Thus we divide our numerical results into two parts, namely the coupled solitons without/with cross-interactions modulations which can be referred to Sec. \ref{sec3a} and Sec. \ref{sec3b}, respectively.

\subsection{Coupled solitons without cross-interactions modulation}
\label{sec3a}

\begin{figure}[tbp]
\begin{center}
\includegraphics[width=1\columnwidth]{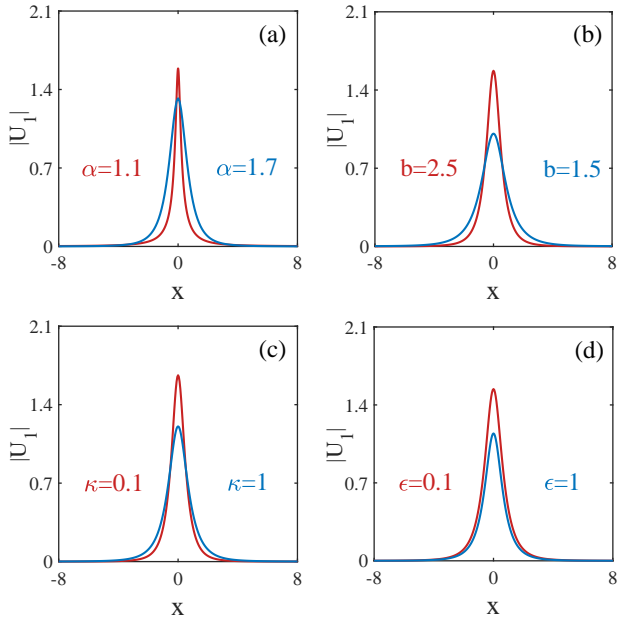}
\end{center}
\caption{Profiles of solitons (shown for components of $U_1$): (a) with different values of $\alpha$ at $\kappa=0.8$, $\epsilon=0.5$, $\gamma=0.001$, $b=2$; (b) with different values of $b$ at $\alpha=1.7$, $\kappa=0.8$, $\epsilon=0.5$, $\gamma=0.001$; (c) with different values of $\kappa$ at $\alpha=1.7$, $\epsilon=0.5$, $\gamma=0.001$, $b=2$; (d) with different values of $\epsilon$ at $\alpha=1.7$, $\kappa=0.8$, $\gamma=0.001$, $b=2$. The profiles of $|U_2|$ are similar to $|U_1|$.}
\label{fig4}
\end{figure}

\begin{figure}[tbp]
\begin{center}
\includegraphics[width=1\columnwidth]{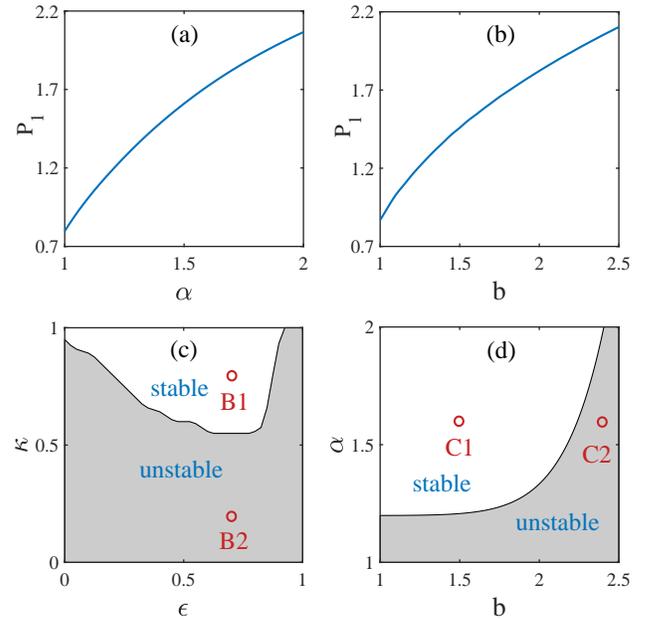}
\end{center}
\caption{(a) $P$ versus $\alpha$ at $\kappa=0.8$, $\epsilon=0.5$, $\gamma=0.001$, $b=2$. (b) $P$ versus $b$ at $\alpha=1.7$, $\kappa=0.8$, $\epsilon=0.5$, $\gamma=0.001$. (c) Stability (white) and instability (gray) domains for the coupled solitons with $\alpha=1.7$, $\gamma=0.001$, $b=1.5$ in the ($\epsilon,\kappa$) plane. The propagations of the solitons corresponding to the points marked by B1 and B2 are shown in Figs. \ref{fig6}(a,b) and \ref{fig6}(c,d), respectively. (d) Stability (white) and instability (gray) domains for the coupled solitons with $\kappa=0.8$, $\epsilon=0.5$, $\gamma=0.001$ in the ($b,\alpha$) plane. The propagations of the solitons corresponding to the points marked by C1 and C2 are displayed in Figs. \ref{fig7}(a,b) and \ref{fig7}(c,d), respectively.}
\label{fig5}
\end{figure}

The $\mathcal{PT}$-symmetric physical model without cross-interactions modulation means that the coefficient $\epsilon=0$, which is a simple model but a nontrivial one. We now display the relevant numerical results in this setting. Note that the numerical results under this setting are shown in Figs. \ref{fig1}$\sim$\ref{fig3}. According to Eq. (\ref{KS}), it is not difficult to find that these real $\mathcal{PT}$-symmetric coupled solutions can only be supported under the condition of $\gamma\leq\kappa$. For simplicity, note also that we use $\mathrm{g}=1$ throughout this article. Profiles of the $\mathcal{PT}$-symmetric coupled solitons with different values of the L\'{e}vy index $\alpha$, different propagation constants  $b$, different linear coupling coefficients $\kappa$, and different values of the gain and loss parameter $\gamma$ are shown in Figs. \ref{fig1}(a$\sim$d), respectively. According to Eqs. (\ref{NLFSES}), we see that the stationary solutions of solitons $U_1(x)$ and $U_2(x)$ have the same real part $U(x)$ and conjugate imaginary parts ${\rm exp}(\pm i\delta/2)$, that is, $U_1(x)=U(x){\rm exp}(-i\delta/2)$, $U_2(x)=U(x){\rm exp}(i\delta/2)$. Note also that the propagation constant $b$ will disappear in the expressions of the stationary solutions of solitons $U_1(x)$ and $U_2(x)$ after we submit Eqs. (\ref{NLFSES}) into Eqs. (\ref{NLFSE}). According to the above expressions of $U_1(x)$ and $U_2(x)$, the profiles of the modulus $|U_2|$ are similar to those of the modulus $|U_1|$, thus we display here only the profiles of $|U_1|$. From Fig. \ref{fig1}(a), the amplitude of coupled solitons decreases and the width increases when $\alpha$ increases. According to Fig. \ref{fig1}(b), the amplitude of coupled solitons increases and the width decreases when $b$ increases. Note also that the amplitude of coupled solitons decreases and the width increases when $\kappa$ increases, as presented in Fig. \ref{fig1}(c), similar to the situation in Fig. \ref{fig1}(a). In Fig. \ref{fig1}(d), the amplitude of coupled solitons increases with the increase of $\gamma$ and the width of solitons decreases with the increase of $\gamma$.

We now focus on the soliton power of the obtained families of coupled solitons, including the relationships of the soliton power $P$ versus the L\'{e}vy index $\alpha$, $P$ versus the propagation constant $b$, and $P$ versus the linear coupling coefficient $\kappa$. We point out that the dependences of the soliton power of the $|U_2|$ component on the  parameters of the model are similar to those of the $|U_1|$ component. Fig. \ref{fig2}(a) shows that the soliton power $P_1$ increases with the increase of $\alpha$, and the increasing speed gradually decreases. According to Fig. \ref{fig2}(b), the soliton power increases when $b$ increases, which shows that the relationship between soliton power and propagation constant satisfies the Vakhitov-Kolokolov (VK) stability criterion $\partial P/\partial b >0$, which is a necessary but not sufficient condition for stable soliton solutions \cite{VK}. The soliton power decreases with the increase of $\kappa$, as presented in Fig. \ref{fig2}(c). The stable and unstable domains of these $\mathcal{PT}$-symmetric coupled solitons are very important, here, Fig. \ref{fig2}(d) displays the stability (white) and instability (gray) domains for the obtained coupled solitons in the ($b,\alpha$) plane. From Fig. \ref{fig2}(d), we can find that the coupled solitons can be stable only when the L\'{e}vy index $\alpha\geq1.2$. The propagation dynamics of the coupled solitons marked by the points A1 and A2 are presented in Figs. \ref{fig3}(a,b) and \ref{fig3}(c,d), respectively. The coupled solitons marked by the point A1 are stable, and both components of the coupled solitons are keeeping their amplitudes and shapes even after a long propagation distance (propagation distance $z=1000$), as displayed in Figs. \ref{fig3}(a,b). On the other hand, the coupled solitons marked by the point A2 are unstable, and both components of the coupled solitons are greately distorted after some propagation distance, as presented in Figs. \ref{fig3}(c,d).

\begin{figure}[tbp]
\begin{center}
\includegraphics[width=1\columnwidth]{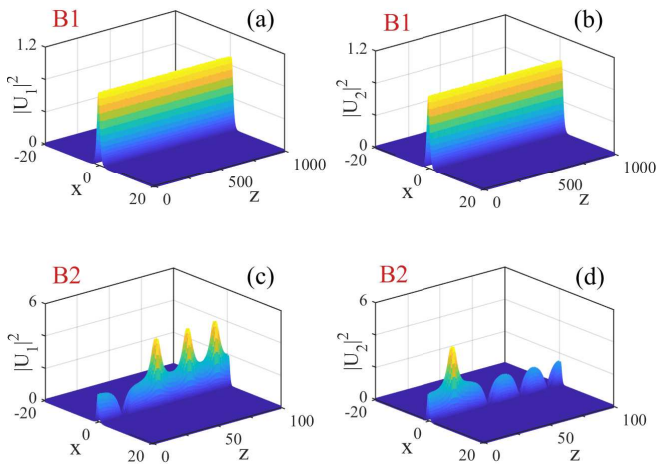}
\end{center}
\caption{Propagations of coupled solitons with different values of $\kappa$ at $\alpha=1.7$, $\epsilon=0.7$, $\gamma=0.001$, $b=1.5$: (a,b) with $\kappa=0.8$; (c,d) with $\kappa=0.2$. The left~/~right column displays the $U_1$~/~$U_2$ component.}
\label{fig6}
\end{figure}

\subsection{Coupled solitons with cross-interactions modulation}
\label{sec3b}

In this subsection, we focus on coupled solitons with cross-interactions modulation, that is, when $\epsilon>0$). It should be mentioned that the numerical results under this setting are displayed in Figs. \ref{fig4}$\sim$\ref{fig9}. Fig. \ref{fig4} shows the profiles of $\mathcal{PT}$-symmetric coupled solitons with different parameters, including the solitons with different L\'{e}vy indices $\alpha$, different propagation constants $b$, different linear coupling coefficients $\kappa$, and different cross-interactions modulations $\epsilon$. The profiles of $|U_2|$ are similar to $|U_1|$, therefore, we only display the profiles of $|U_1|$ in this subsection.

The amplitude of coupled solitons decreases and the width of increases when $\alpha$ increases, as shown in Fig. \ref{fig4}(a). In Fig. \ref{fig4}(b), the amplitude of coupled solitons increases while the width decreases with the increase of $b$. According to Fig. \ref{fig4}(c), the amplitude of coupled solitons decreases and the width increases when $\kappa$ increases. From Fig. \ref{fig4}(d), the amplitude and width of coupled solitons increase when $\epsilon$ increases. Obviously, the results shown in Figs. \ref{fig4}(a$\sim$c) are similar to their counterparts shown in Figs. \ref{fig1}(a$\sim$c) (without cross-interactions modulation).

Fig. \ref{fig5} displays the soliton power and relevant stability area of these $\mathcal{PT}$-symmetric coupled solitons with cross-interactions modulation. Here, the soliton power of $|U_2|$ is similar to that of $|U_1|$. Fig. \ref{fig5}(a) demonstrates that the soliton power $P_1$ increases with the increase of $\alpha$, and the growth speed gradually diminishes. From Fig. \ref{fig5}(b), the soliton power increases when $b$ increases, which demonstrates that the relationship between soliton power and propagation constant for the obtained coupled solitons also satisfies the VK stability criterion. The stable and unstable domains of the coupled solitons with cross-interactions modulation are also very important. Here, Figs. \ref{fig5}(c,d) show the stability (white) and instability (gray) domains for the coupled solitons in the ($\epsilon,\kappa$) and ($b,\alpha$) planes, respectively. According to Fig. \ref{fig5}(c), the coupled solitons can be stable only when $\kappa$ is above a certain threshold, which is about 0.55 in this case. Fig. \ref{fig5}(d) demonstrates that these coupled solitons can be stable only when $\alpha\geq1.2$. Further, by comparing the results of Fig. \ref{fig5}(d) and Fig. \ref{fig2}(d), we can clearly see that the stable domain of Fig. \ref{fig5}(d) is much broader than the one in Fig. \ref{fig2}(d). Note that other parameters in Fig. \ref{fig5}(d) and Fig. \ref{fig2}(d) are the same except for cross-interactions modulation $\epsilon$, therefore, we can draw a conclusion that $\epsilon$ can greatly affect the stability and instability domains.

\begin{figure}[tbp]
\begin{center}
\includegraphics[width=1\columnwidth]{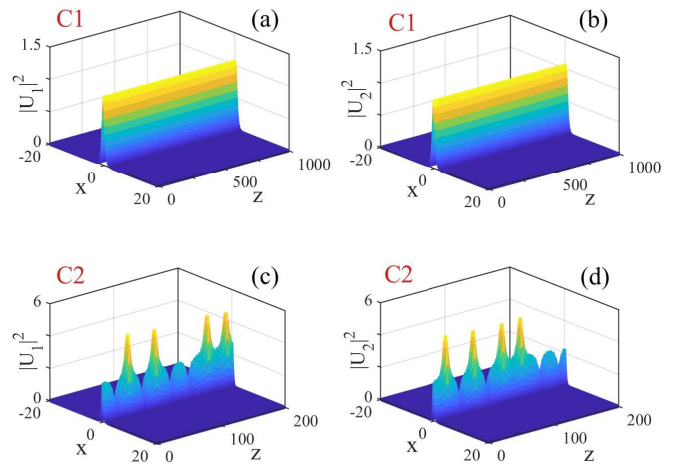}
\end{center}
\caption{Propagations of coupled solitons with different values of $b$ at $\alpha=1.6$, $\kappa=0.8$, $\epsilon=0.5$, $\gamma=0.001$: (a,b) with $b=1.5$; (c,d) with $b=2.4$. The left~/~right column displays the $U_1$~/~$U_2$ component.}
\label{fig7}
\end{figure}

The propagation dynamics of the coupled solitons marked by the points (B1,B2) [in Fig. \ref{fig5}(c)] and (C1,C2) [in Fig. \ref{fig5}(d)] are presented in Figs. \ref{fig6} and \ref{fig7}, respectively. The coupled solitons marked by the points B1 and C1 are stable, and both components of these coupled solitons are keeping their amplitudes and shapes even after long-distance propagations (after a propagation distance $z=1000$), as displayed in Figs. \ref{fig6}(a,b) and \ref{fig7}(a,b). On the other hand, the coupled solitons marked by the points B2 and C2 are unstable, and both components of these coupled solitons are greately distorted even after short-distance propagations, as presented in Figs. \ref{fig6}(c,d) and \ref{fig7}(c,d).

\begin{figure}[tbp]
\begin{center}
\includegraphics[width=1\columnwidth]{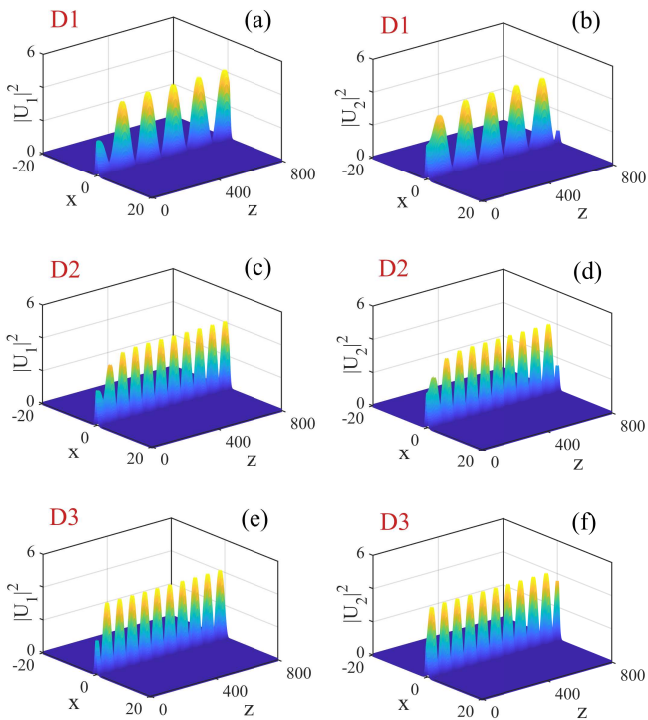}
\end{center}
\caption{Stable propagations of oscillating solitons at $\alpha=1.8$, $\epsilon=1$, $b=2$: (a,b) with $\kappa=0.02$, $\gamma=0.01$; (c,d) with $\kappa=0.04$, $\gamma=0.01$; (e,f) with $\kappa=0.04$, $\gamma=0.038$. The left~/~right column displays the $U_1$~/~$U_2$ component. Note that the oscillating period can be controlled only by varying the value of $\kappa$. The soliton power $P$ versus propagation distant $z$ for panels (a)$\sim$(c) are displayed in Figs. \ref{fig9}(a)$\sim$\ref{fig9}(c), respectively.}
\label{fig8}
\end{figure}

\begin{figure}[tbp]
\begin{center}
\includegraphics[width=1\columnwidth]{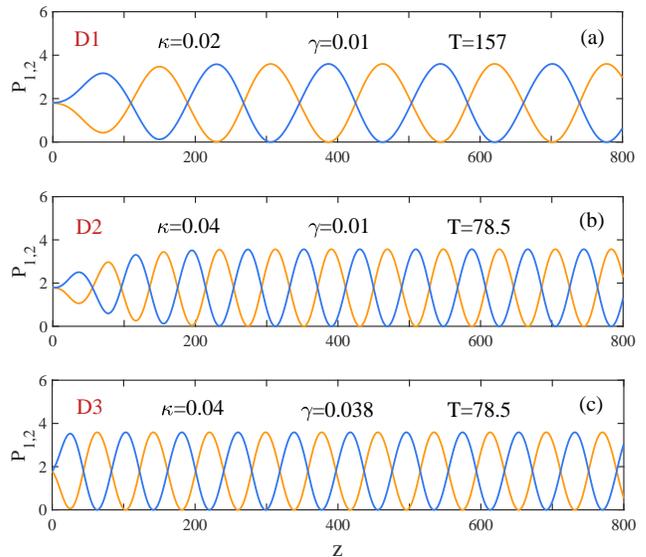}
\end{center}
\caption{Soliton power $P$ versus $z$ for the propagation displayed in Figs. \ref{fig8}(a)$\sim$\ref{fig8}(c). The yellow and blue lines stand for $P_1$ and $P_2$, respectively. Note that the parameter $\gamma$ can not affect the oscillating period ${\rm T}$, while it can affect the convergence speed to the equilibrium state.}
\label{fig9}
\end{figure}

Next we turn to investigate the formation and propagation properties of the oscillating $\mathcal{PT}$-symmetric coupled solitons in the model of cross-interactions modulation. It was mentioned before that $\mathrm{g}=1$ is used throughout this paper; here we find that the unstable coupled solitons become oscillating coupled solitons when $\epsilon=1$. Some examples of the oscillating coupled solitons are displayed in Fig. \ref{fig8}, which shows the propagation dynamics of $U_1$ in the left column and the corresponding propagation dynamics of $U_2$ in the right column. Figs. \ref{fig8}(a,b) display the propagation dynamics of the oscillating coupled solitons with a long oscillating period, and Figs. \ref{fig8}(c$\sim$f) present the propagation dynamics of the oscillating coupled solitons with a short oscillating period. According to Figs. \ref{fig8}(a,b), we see that when the amplitude of one component decreases, the amplitude of the other component would increase at the same time, which is related to the coefficient of gain and loss $\gamma$.

To further study the properties of these interesting oscillating $\mathcal{PT}$-symmetric coupled solitons, we draw the relationships between their soliton power $P$ and propagation distance $z$ in Fig. \ref{fig9} (whose propagation dynamics are presented in Fig. \ref{fig8}). In Fig. \ref{fig9}, the yellow and blue lines stand for P1 and P2, respectively. By comparing the results in Figs. \ref{fig9}(a,b), we can see that the oscillating period depends on the linear coupling coefficient $\kappa$, and it decreases when $\kappa$ increases. In addition, we point out that the convergence speed of transforming the coupled solitons into oscillating coupled solitons is strongly related to the specific value of the gain/loss parameter $\gamma$. That is to say, the larger $\gamma$ is, the faster the coupled solitons transform into the oscillating coupled solitons. It should be emphasized that we have verified the above conclusions by using different sets of parameters. Furthermore, we find that these oscillating $\mathcal{PT}$-symmetric coupled solitons can be supported under the condition of $\mathrm{g}=\epsilon>0$, which is not limited to $\mathrm{g}=\epsilon=1$.

\section{Conclusion}
\label{sec4}
In this paper, we have investigated the existence and propagation dynamics of the families of coupled solitons in $\mathcal{PT}$-symmetric physical models with gain and loss in fractional dimension, including the coupled models with and without cross-interactions modulation. The profiles, powers, stability domains, and propagation properties of these $\mathcal{PT}$-symmetric coupled solitons are studied in both physical settings. According to our study, the profiles of such $\mathcal{PT}$-symmetric coupled solitons are sensitive to the values of propagation constant, and the stable coupled solitons can only exist under the condition that the L\'{e}vy index is greater than a critical value. In addition, the obtained results  reveal that the cross-interactions modulation can greatly affect the stability and instability domains of the coupled solitons. In other words, the stability area of the model with cross-interactions modulation is much broader than the one without cross-interactions modulation. Further, we have also studied the oscillating $\mathcal{PT}$-symmetric coupled solitons that can be supported when the coefficients of the self- and cross-interactions modulations are the same. Besides, the oscillating period of these coupled solitons can be controlled by varying the value of the linear coupling coefficient, and the convergence speed of coupled solitons that are transforming into oscillating coupled solitons is only related to the value of the gain/loss parameter.

The physical model studied in this work can be extended to the (2+1)-dimensional setting, which is an interesting but a quite difficult problem due to the presence of the critical collapse \cite{COLLAPSE}. Also, the introduction of linear periodic potentials into the physical model is another interesting issue to be investigated, because the coupled gap solitons \cite{CGS} can form in that setting.

\section*{Acknowledgment}
We thank Dumitru Mihalache and Jiawei Li for their help on improving the writing of the manuscript.

\section*{Funding}
This work was supported by National Major Instruments and Equipment Development Project of National Natural Science Foundation of China (No. 61827815), by National Natural Science Foundation of China (No. 62075138), and by Science and Technology Project of Shenzhen (Nos. JCYJ20190808121817100, JCYJ20190808164007485).

\section*{Conflict of interest}
\label{sec5}
The authors declare that they have no conflict of interest.

\end{document}